# A Novel Chaotic Uniform Quantizer for Speech Coding

Osama A. S. Alkishriwo[1*]

[1] alkishriewo@yahoo.com

[1] Department of Electrical and Electronic Eng., College of Eng., University of Tripoli, Libya

*Corresponding author email: alkishriewo@yahoo.com

**ABSTRACT**

Quantization is an essential step in the analog-to-digital conversion process and it is very important in all modern telecommunication systems. In this paper, a novel chaotic uniform quantizer is proposed and its application for speech coding is presented. The proposed system consists of three stages: two PCM coders separated by an XOR operation with a chaotic sequence, where the first step is used for continuous signal sampling and second stage performs data encryption, while the third stage provides additional data compression. The performance of the presented quantizer for Laplacian distributed signals and real speech signals is investigated and compared with that of the well-known uniform and non-uniform quantizers. Simulation results show that the proposed quantizer provides secured data with higher levels of SQNR compared to others.

**Keywords:** Quantization; Uniform quantization; Non-uniform quantization; Source coding; Encryption; Chaotic systems.

## 1  Introduction

In all modern telecommunication systems, the analog–to–digital conversion is very important phase in the way of processing analog signals. It consists of two main steps which are quantization and coding. Quantization provides a means to represent signals efficiently with acceptable fidelity for signal compression, while coding decides exactly which code–words to assign to each distinct quantization level [1].

Existing quantization schemes can be classified into two categories, namely, uniform quantization and nonuniform quantization [2]. Uniform quantization is widely used due to its simplicity of implementation, but not optimal for signals with nonuniform distribution in terms of mean square error. While nonuniform quantization is much more complex, it is in general causes less information loss than uniform quantization, especially for small quantization resolutions. Lloyd–Max quantization is a major type of nonuniform quantization [3, 4]. It is optimal in the sense of mean squared error (MSE), but it is computationally intensive. Companding, which consists of nonlinear transformation and uniform quantization, is a technique capable of trading off quantization performance with complexity for nonuniform quantization [5].

In literature, many nonuniform quantizers have been developed to meet the challenge of designing a low complexity and high signal–to–quantization noise ratio (SQNR). In [6, 7] sophisticated compression models based on fixed–rate scalar quantizer for Laplacian probability density function have been recently proposed. The problem of support region optimization has been extensively considered in the field of scalar quantization [8]. The optimization of the support region of the product polar companded quantizer is presented in [9]. This resulted in SQNR increase, but in a more complex encoding/decoding procedure. Although a great number of quantizers have been developed to provide an additional enhancement of the quantized signal quality, there is still a need to continue the research in this field.

In this paper, a novel chaotic uniform quantizer is proposed. It consists of three stages: two PCM coders separated by an XOR operation with a chaotic sequence. The first step is used for continuous signal sampling using rough quantization with a large number of quantization levels. After that, encoded data are XORed with a uniformly distributed random sequence which is generated from chaotic dynamic system to perform data encryption. The encrypted data are decoded to obtain discrete samples, which are further quantized using a quantizer with a small number of quantization levels in order to provide additional compression. The proposed quantizer is evaluated by means of a computer simulation using synthetic Laplacian source signals and real speech signals. The presented quantizer provides gain in the signal to quantisation noise ratio, encryption of the quantized samples, and compression over the conventional uniform quantizer as well as companding quantizer. This indicates the possibility of practical application of the chaotic proposed quantizer in the contemporary transmission of speech signals.

The rest of the paper is organized as below. Section II presents the preliminaries of chaotic uniform quantizer. Section III describes the proposed chaotic uniform quantizer. Simulation results, comparison and discussion are given in Section IV. Finally, conclusions are summarized in Section V.

## 2 Preliminaries of Chaotic Uniform Quantizer

### 2.1 Chaotic Dynamic System

Chaos systems are considered suitable for practical use because of its complex dynamic behaviors. They provide a good combination of speed and high security. They have many excellent intrinsic properties, such as high sensitivity to initial conditions and control parameters, which are the desired properties in the application of encryption. The three dimensional discrete chaotic system, which are presented in [10], is defined as follows

$$x_{n+1} = \left[\frac{\alpha \times (x_n - x_n^2)}{(y_n - y_n^2)}\right] \ mod \ 1$$

$$y_{n+1} = \left[\frac{\beta \times (y_n - y_n^2)}{(z_n - z_n^2)}\right] \ mod \ 1 \quad (1)$$

$$z_{n+1} = \left[\frac{\gamma \times (z_n - z_n^2)}{(x_n - x_n^2)}\right] \ mod \ 1$$

The chaotic behaviour of the system is observed when the control parameters are chosen as $0.5 < \alpha, \beta, \gamma < 4$ with the initial conditions $x_n, y_n$, and $z_n \in [0, 1]$.

**2.2   Uniform Quantizer**

A uniform quantizer splits the mapped input signal into quantization steps of equal size. If $X$ is a random variable with the probability density function $p_X(x)$ at input of the quantizer is converted to one of $Q$ allowable levels, $m_1, m_2, ..., m_Q$ and $Y$ is a discrete random variable at output of the quantizer. Then, the quantizer $q$ maps $X$ to $Y$ as follows:

$$Y = q(X) = m_i, \quad i = 1, 2, 3, ..., Q \quad (2)$$

Thus the quantization error, $e_q = x - y$, is a random variable with pdf $p_{e_q}(e_q)$ and the average distortion is,

$$D = E\{e_q^2\} = \int_{-\infty}^{+\infty} (X - Y)^2 \ p_X(x) \ dx \quad (3)$$

where $E\{.\}$ is the expectation value. The signal to quantization noise ratio ($SQNR$) is obtained by dividing the input source variance $\sigma^2$ over the average distortion ($D$) as follows:

$$SQNR = 10 \ \log_{10}\left(\frac{\sigma^2}{D}\right) \quad (4)$$

**3   Proposed Chaotic Uniform Quantizer**

The configuration of the proposed chaotic uniform quantizer is shown in Figure 1. The quantization process can be achieved using three main steps. In the first step, analog-to-digital (A/D) conversion is performed using a quantizer with a high number of quantization levels $Q_1$, whose purpose is to convert analog signal to discrete samples. Then, the quantized samples are encoded and XORed with a uniform random sequence generated from chaotic system. The resulted data are decoded to obtain an encrypted discrete time signal which has a uniform probability density function. The aim of the second stage is to provide additional signal compression by using a low number of quantization levels $Q_2$ where ($Q_2 < Q_1$). The key stream in the chaotic system is composed of control parameters $\alpha, \beta, \gamma$ and initial values $x_0, y_0, z_0$.

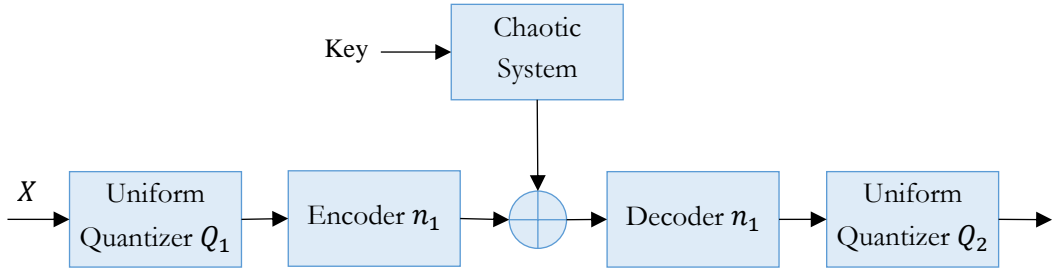

**Figure 1:** *Proposed chaotic uniform quantizer.*

## 4   Simulation Results

The input–output characteristics of uniform quantizer, nonuniform quantizer, and chaotic quantizer are shown in Figure 2. It is known that uniform quantizer has fixed step size and fixed output level as given in Figure 2(a), while nonuniform quantizer has variable step size and fixed output level as illustrated in Figure 2(b). However, chaotic quantizer has fixed step size and variable output level as shown in Figure 2(c).

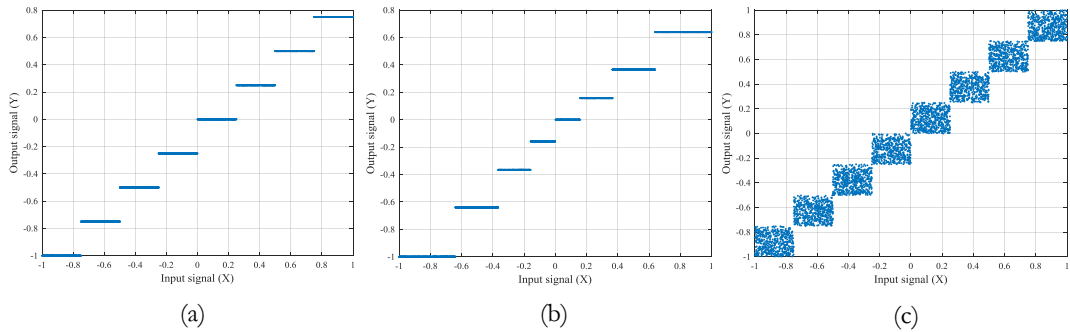

**Figure 2:** *Quantizer characteristics : (a) Uniform quantizer, (b) Nonuniform quantizer, (c) Chaotic quantizer.*

The performance of the proposed chaotic quantizer in quantization of signals having Laplacian probability density function is shown in Figure 3(a). The choice of Laplacian distribution is made so as to match the data typically found in speech coding problem. The results of chaotic quantizer are compared with the results presented by conventional uniform and nonuniform quantizers. As a direct application of the proposed chaotic quantizer for speech coding, it has been applied to a speech signal and the results are reported in Figure 3(b). In both cases, the performance of chaotic quantizer in terms of SQNR is superior to the performances of traditional uniform and nonuniform quantizers. For instance, for the case of Laplacian source, the chaotic quantizer produces $3\ dB$ and $1\ dB$ SQNR on average higher than uniform and nonuniform quantizers, respectively. Similarly, for the case of speech signal the improvement on the values of SQNR are $2.5\ dB$ and $0.8\ dB$.

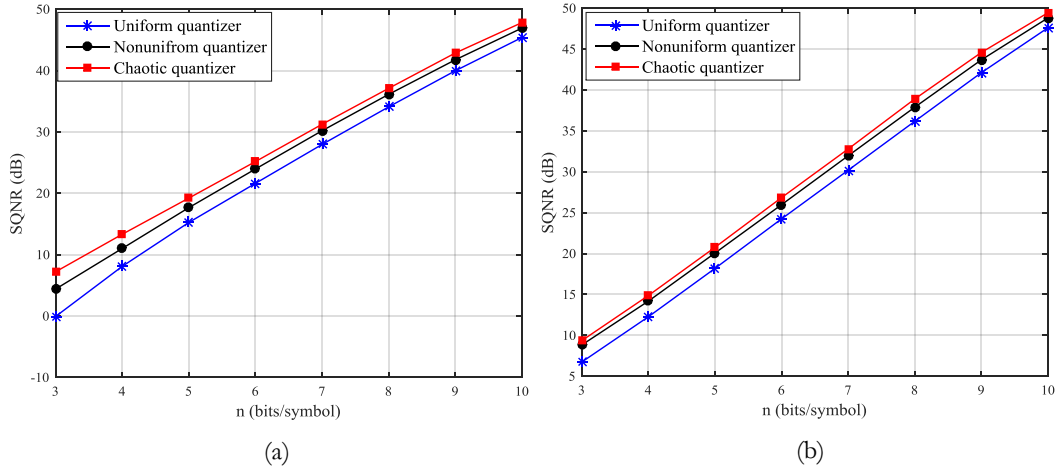

**Figure 2:** *Signal to quantization noise ratio versus number obits per symbol for, (a) Source with Laplacian probability density function, (b) Speech signal.*

An important advantage of the proposed chaotic quantizer is the quantization of samples and encrypting them at the same time. The used chaotic system given in (1) has three initial values and three control parameters. Thus, if the precision is set to be $10^{-15}$, the key space can reach $10^{6\times15} \approx 2^{299}$, which can efficiently resist the brute-force attack.

## 5    Conclusions

In this paper, the chaotic uniform quantizer is proposed and its performance for input signal with Laplacian distribution and its application for speech signal processing are explored. Experimental results demonstrate that the chaotic uniform quantizer is superior to the conventional uniform and nonuniform quantizers at all bit per symbol region. It has been shown that the proposed chaotic quantizer provides more constant and higher level of SQNR, which can be considered via the gains in the SQNR that range up to $3\ dB$ as illustrated in the results section. Finally, the quantized samples are encrypted with a chaotic sequence which has a key space of $2^{299}$.

## References

[1]  N. S. Jayant and P. Noll,  *Digital Coding of Waveforms*,  Prentice Hall, Upper Saddle River, N. J., 1984.
[2]  R. M. Gray and D. L. Neuhoff, "Quantization," *IEEE Transactions on Information Theory*, vol. 44, no. 6, pp. 2325-2383, Oct. 1998.
[3]  S. P. Lloyd, '"Least squares quantization in PCM," *IEEE Transactions on Information Theory*, vol. IT-28, no. 2,  pp. 129-137, Mar. 1982.
[4]  J. Max, "Quantizing for minimum distortion," *IRE Transactions on Information Theory*, vol. IT-6, pp. 7-12, Mar. 1960.

[5] ITU-T, Recommendation G.711, Pulse Code Modulation (PCM) of Voice Frequencies, International Telecommunication Union, 1972.
[6] Z. Peric and J. Nikolic, "An adaptive waveform coding algorithm and its application in speech coding," *Digital Signal Processing*, vol. 22, no. 1, pp. 199-209, Jan. 2012.
[7] Z. Peric and J. Nikolic, "High--quality Laplacian source quantization using a combination of restricted and unrestricted logarithmic quantizers," *IET Signal Processing*, vol. 6, no. 7, pp. 633-640, Nov. 2012.
[8] J. Nikolic, Z. Peric, and A. Jovanovic, "Two forward adaptive dual--mode companding scalar quantizers for Gaussian source," *Signal Processing*, vol. 120, no. 3, pp. 129-140, Mar. 2016.
[9] Z. Peric, M. D. Petkovic, J. Nikolic, and A. Jovanovic, "Support region estimation of the product polar companded quantizer for Gaussian source," *Signal Processing*, vol. 143, pp. 140-145, Feb. 2018.
[10] M. Y. Valandar, P. Ayubi, M.J. Barani, "A new transform domain steganography based on modified logistic chaotic map for color images," *Journal of Information Security and Applications*, vol. 34, no. 2, pp. 142-151, Jun. 2017.